\documentclass[prl,showpacs,twocolumn,superscriptaddress]{revtex4-1}
\usepackage{graphicx,amssymb,amsmath,bm,color,bbm}
\newcommand{\NN}{L}
\newcommand{\jav}[1]{{#1}}

\begin{document}
\definecolor{darkgreen}{rgb}{0,0.5,0}
\definecolor{matlabmagenta}{rgb}{1,0,1}

\title{Quantum spin-Hall insulator in quantized electromagnetic field: Dicke-type quantum phase transition}
\title{From Floquet to Dicke: Quantum spin-Hall insulator in quantized electromagnetic field}
\title{From Floquet to Dicke: quantum spin-Hall insulator interacting with quantum light}

\author{Bal\'azs Gul\'acsi}
\affiliation{Department of Physics and BME-MTA Exotic  Quantum  Phases Research Group, Budapest University of Technology and
  Economics, 1521 Budapest, Hungary}
\author{Bal\'azs D\'ora}
\email{dora@eik.bme.hu}
\affiliation{Department of Physics and BME-MTA Exotic  Quantum  Phases Research Group, Budapest University of Technology and
  Economics, 1521 Budapest, Hungary}

\date{\today}

\begin{abstract}
Time-periodic perturbations due to classical electromagnetic fields are useful to engineer 
the topological properties of matter using the Floquet theory.
Here we investigate the effect of \emph{quantized} electromagnetic fields by focusing 
on the quantized light-matter interaction on the edge state of a  quantum spin-Hall insulator.
A Dicke-type superradiant phase transition occurs at arbitrary weak coupling,  the electronic spectrum acquires a finite gap and the resulting ground state manifold
is topological with Chern number $\pm 1$.
When the total number of excitations is conserved, a photocurrent is generated along the edge, being pseudo-quantized 
as $\omega\ln(1/\omega)$ in the low frequency limit,
and decaying as $1/\omega$ for high frequencies with $\omega$ the photon frequency.
The photon spectral function exhibits a clean Goldstone mode, a Higgs like collective mode at the optical gap and the polariton continuum.
\end{abstract}

\maketitle

\emph{Introduction.}
The conventional band theory of solids has attracted renewed attention after the theoretical prediction and experimental 
discovery of topological insulators (TIs)\cite{hasankane,qi,kitaev}.
The topological protection of their edge/surface states, together with the spin-momentum locking due to the strong spin orbit coupling, yields a variety of novel phenomena,
such as magnetoelectric effect, axion electrodynamics, surface Hall effect, not to mention topological superconductivity and Majorana fermions.
In addition to their theoretical appeal, they also hold the promise to realizing practical quantum computers and  spintronical devices.

In spite of the growing interest, materials with non-trivial topological properties are still scarce. 
Achieving a topological band structure often requires engineering  the Bloch band structure.
Recently, time periodic perturbations, mostly due to electromagnetic fields,  have been proposed and used\cite{lindner,cayssol,rechtsman,gedik,asboth} 
to manipulate the band structure using the Floquet theory, the temporal analogue of Bloch states.
Although the steady state of the resulting Floquet topological insulators often possesses a topology different  from that of their static parents,
the actual occupation of these states is still far from being understood\cite{dehghani} albeit
it is an essential ingredient for determining the physical response of the system.  
The occupation depends on the sources of relaxation, e.g. coupling to heat baths,
momentum scattering, interaction etc., and the conventional ground-state picture with the minimal energy configuration cannot be  used for these driven models.
To circumvent this problem, a classical driving field can be replaced by its quantum counterparts having its own quantum dynamics.
This idea has been pursued for the creation of Majorana fermions using not only classical driven systems\cite{zoller} but rather their 
quantum counterparts using high-Q electromagnetic cavity\cite{trif}.

The properties of classical and quantized electromagnetic fields have long been investigated.  In connection with their interaction with a 
single atom\cite{gerry,aravind}, e.g., many similarities have been demonstrated  between the corresponding Rabi and Jaynes-Cummings 
models. Besides the similarities, quantum fields also provide novel effects and collective phenomena such as spontaneous symmetry breaking and the 
superradiant phase  transition\cite{gerry,wanghioe,emarybrandes} due to light-matter interaction in the Dicke model\cite{nagy,baumann}.

{In this work,  we  combine  TIs and quantized electromagnetic fields to explore the quantum mechanical version of the model of Ref.~\cite{dorafloquet}, 
shown to display a Floquet topological phase transition. We demonstrate  that
 a superradiant quantum phase transition    exists in the quantum context,  and is   accompanied by a  transition to a topologically 
 non-trivial ground state manifold  and by an induced photocurrent.}

\begin{figure}[h!]
\centering
\includegraphics[width=5cm]{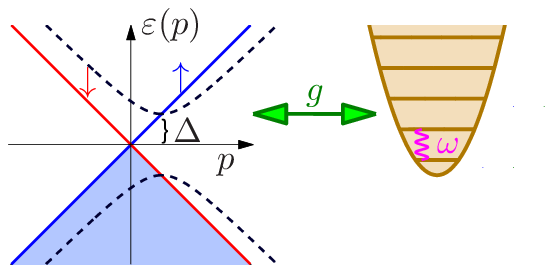}
\includegraphics[width=3.5cm]{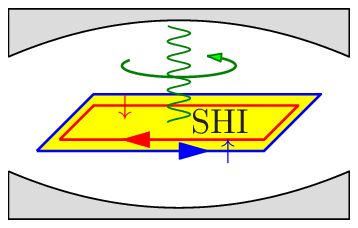}
\caption{(Color online) Left: a single $\omega$ frequency boson mode is coupled to a continuum of fermionic degrees of freedom, which is
a quantum spin-Hall insulator in the present case. This is the essential difference between our setup and conventional Dicke/Tavis-Cummings models, where the fermions
possess a discrete spectrum. The new fermionic spectrum becomes gapped and shifted (dashed line).
Right: cartoon of the experimental setup corresponding to Eq. \eqref{hamilton}   of SHI with spin filtered edge states placed into an optical cavity.}
\label{bosoncontinuum}
\end{figure}

\emph{The model.} 
\jav{We consider the edge  dynamics of a spin-Hall insulator\cite{kanemele1} (SHI), interacting with
circularly polarized quantized electromagnetic field through the Zeeman term (see Fig. \ref{bosoncontinuum}).}
The Hamiltonian  is
\begin{gather}
H=\omega a^+a+v\sum_p pS^z_p+\frac{g}{\sqrt \NN}\sum_p\left(a^+S^-_p+aS^+_p\right),
\label{hamilton}
\end{gather}
Here $v$ is the Fermi velocity of the edge state,
 $\omega$ and $a$ denote  the energy and annihilation operator of the cavity mode,  and $c_{p,\sigma}$ annihilates
an electron with momentum $p$ and spin $\sigma$. 
The last term describes   the Zeeman interaction with the spin operators expressed as
$S^+_p=c^+_{p,+}c_{p,-}$ and $S^z_p=\sum_{\sigma=\pm}\sigma c^+_{p,\sigma}c_{p,\sigma}$.  The Zeeman coupling
depends on the photon's frequency \cite{gerry} as $g=\sqrt{ \tilde g|\omega|}$, 
{and $\NN$ is the dimensionless length of the edge} 
\footnote{{The coupling $g/\sqrt{\NN}$ is inversely proportional to the cavity's volume, $\sqrt{V}$}, and turns out to be small for usual 
radio frequency resonators.}.

{ Let us first highlight} 
the connection of Eq. \eqref{hamilton} to Floquet topological insulators using the one body picture of the polaritons.
{Considering just a single momentum} $p$, the Hamiltonian reduces to the Jaynes-Cummings model\cite{gerry}.
The spectrum consist of energy pairs 
 $\epsilon_{\pm}(p,m)=\omega(m-\frac 12)\pm\sqrt{(vp-\frac\omega 2)^2+\frac{g^2}{\NN}m}$ with $m$ being a positive integer, 
plus a single mode $\epsilon_{-}(p,0)=-vp$ 
\footnote{Note that its wavefunction is also an eigenstate of the many-body Hamiltonian as \unexpanded{$|0\rangle\otimes\prod_p|\downarrow_p\rangle $} with eigenenergy 0, 
describing a fully polarized state that carries an electric current $j=eW/\pi$}. 
These polariton energies are closely related to the Floquet spectrum of the same problem in the presence of classical electromagnetic field \cite{cayssol},
when the photon state is prepared in a coherent state with $m\gg 1$. Importantly, similar to the classical case, 
the spectrum becomes gapped due to interaction with the quantized light field for all $m>0$~\cite{kibis}, and for 
each  fixed $m>0$,  the contribution of all    
  $\epsilon_{-}(p,m)$ states amounts  in an  electric current  identical to that in the classical  case, 
  quantized as $j=\frac{e\omega}{2\pi}$ in agreement with  Floquet 
adiabatic charge pumping.

This one body description {has limited meaning, since} all spins interact simultaneously 
with the same bosonic mode, and the many-body spectrum involves all of them.
The  Hamiltonian  Eq.~\eqref{hamilton} is, however, integrable\cite{dukelsky}  for any  
number of spins, and coincides with the inhomogeneous Dicke model~\cite{Tsyplyatyev}. 
In addition to the  energy, the total number of excitations, $\mathcal Q=a^+a+\frac 12\sum_p S^z_p$ is a conserved quantity, which
generates a $U(1)$ symmetry. Without $g$, the photon state is unpopulated and the total spin is unpolarized, therefore $\langle \mathcal Q\rangle=0$.
Note that this applies to positive helicity ($\Lambda$) of light, \jav{$\Lambda=1$}, 
for negative helicity \jav{($\Lambda=-1$)} the replacement $a\leftrightarrow a^+$ should be made in the Zeeman term, 
and $a^+a-\frac 12\sum_p S^z_p$ is conserved.

\emph{Mean field theory.} 
{In the thermodynamic limit $\NN\rightarrow\infty$ the bosonic state becomes macroscopically populated, and the mean field description 
$\langle a\rangle\approx\sqrt{n\NN}\exp(i\phi)$ becomes exact~\cite{wanghioe,emarybrandes}.}
The $U(1)$ symmetry of the normal phase is {spontaneously broken and the bosonic field acquires a nonzero, macroscopic mean value, implying 
superradiance.}
Note that integrating out the photon field in Eq. \eqref{hamilton} leads to an effective long range electron-electron interaction
of the form $-\tilde g \sum_{p,k}S^+_pS^-_k$. In contrast to the  short range interactions in low dimensional models, this  long-ranged interaction 
 facilitates the  $U(1)$ symmetry breaking even at finite temperatures.


\jav{The ground state is determined by either choosing the absolute minimum energy configuration\cite{wanghioe,emarybrandes}
from all possible $\mathcal Q$ sectors, referred to as \emph{unconstrained} solution, or by
fixing the total number of excitations\cite{eastham,keeling,zhangdicke}
in
 a \emph{constrained} solution (i.e. grand canonical ensemble).
The former situation is achievable in the presence of coupling to external noise or dissipation, inducing transition between various $\mathcal Q$ sectors,
while the constrained solution corresponds to a situation in which polariton
decay, arising from coupling the cavity mode or  the electronic excitations to other modes,
is slow
compared with the time required to reach thermal equilibrium at a
fixed polariton number.
 The $\mathcal Q=0$ constraint  is taken into account by adding $-\lambda \mathcal Q$ to the Hamiltonian with $\lambda$ the Lagrange multiplier\cite{eastham,keeling}.}



Within  mean field theory, the fermionic excitation spectrum reads as
\begin{gather}
E_\alpha(p)=\alpha\sqrt{\left(vp-\frac\lambda 2\right)^2+\Delta^2},
\label{elspec}
\end{gather}
where $\alpha=\pm$ and, similarly to the case of classical light\cite{cayssol} and to the one body picture,  a gap $\Delta=g\sqrt n$ 
opens in the spectrum due to the the quantized light -- matter interaction.
The mean field parameters, $n$, $\phi$ and $\lambda$ are determined by minimizing the ground states energy
\begin{gather}
\frac{E_0}{\NN}=(\omega-\lambda)n-\frac{\rho}{2}\int\limits_{-W}^{W}d\varepsilon \sqrt{\left(\varepsilon-\frac\lambda 2\right)^2+g^2n}.
\end{gather}
where {$\rho=1/\pi v$ is the 1D density of states and $W$ is a high energy cutoff. }
The mean field equations, ${\partial E_0}/{\partial n}=0$ and ${\partial E_0}/{\partial \lambda}=0$ 
yield in the weak coupling limit
\begin{gather}
 \omega-\lambda=\frac{\rho g^2}{2}\ln\left(\frac{2W}{g\sqrt{n}}\right),\hspace*{6mm} \lambda=-\frac{4n}{\rho},
\label{mfeqs}
\end{gather}
respectively.
Both equations are to be solved in the constrained case, while the unconstrained case requires only the first one in Eq. \eqref{mfeqs} with $\lambda=0$ and the second equation should be ignored.

The phase of the order parameter, $\phi$ remains undetermined {in both cases}
 since the ground state energy shows a Mexican hat in the space of Re$\langle a\rangle$ and
Im$\langle a\rangle$.  This degree of freedom  corresponds to
 a gapless Goldstone mode from the broken $U(1)$ symmetry
in the superradiant phase of Eq. \eqref{hamilton}, similar to the Tavis-Cummings model\cite{xiang}.
By tuning the phase $\phi$ adiabatically from 0 to $2\pi$, one sweeps through the degenerate ground state manifold with no energy cost.
The $(p,\phi)$ space can be mapped to the unit sphere with 
${\bf d}_{\alpha}=(\langle S^x_p\rangle,\langle S^y_p\rangle,\langle S^z_p\rangle)=(\Delta\cos\phi,\Delta\sin\phi,vp-\lambda/2)/E_\alpha(p)$.
The winding number of the mapping defines the Chern number, which reads for band $\alpha$ \jav{and given helicity of light $\Lambda=\pm 1$} as 
\begin{gather}
\mathcal{C}_\alpha=\int\limits_{-\infty}^{\infty}dp\int\limits_0^{2\pi} \frac{d\phi}{4\pi}{\bf d}_{\alpha} \cdot
  \left(\partial_p{\bf  d}_\alpha\times \partial_\phi {\bf  d}_\alpha\right)=-\alpha~\jav{\Lambda}.
\label{chern}
\end{gather}
The $\mathcal{C}_\alpha$ is analogous to the Chern number of the Floquet case\cite{dorafloquet}. 
The superradiant ground state manifold is thus {topologically non-trivial}, irrespective of the value of $\lambda$.

We emphasize again that  the phase degree of freedom from $U(1)$ symmetry breaking in a 1D fermionic model arises due to the photon mediated infinitely
 long range interaction between electrons, while artificially created phases define similar Chern numbers in XY spin chain\cite{yqma}.
A photocurrent can also generated along the edge as $\langle j\rangle =-{e\lambda}/{2\pi}$, depending on the value of $\lambda$.

Not only the bosonic field develops a macroscopic population, but the time reversal symmetry is also
broken, which is indicated by the finite electronic spin density perpendicular to the $z$ direction as
\begin{gather}
\langle S^{\pm}\rangle=-\exp(\mp i\phi)\frac{\rho \Delta}{2}\ln\left(\frac{2W}{\Delta}\right).
\end{gather}

\emph{Unconstrained solution.}
The ground state is the absolute minimum energy configuration over all possible $\mathcal Q$ sectors as in  Refs. \cite{wanghioe,emarybrandes}, since
the total excitation number is not fixed.
The order parameter follows a typical weak-coupling, BCS-like form as
\begin{gather}
n=\left(\frac{2W}{g}\right)^2\exp\left(-\frac{4\omega}{\rho g^2}\right)=\frac{4W^2}{\tilde g \omega}\exp\left(-\frac{4}{\rho \tilde g}\right),
\label{spsol1}
\end{gather}
where the $\omega$ dependence is made explicit, and the gap, $\Delta=2W\exp(-2/\rho\tilde g)$ is independent of $\omega$. 
Since the electrons feel the photon mode through the opening of the gap, which is frequency independent,
the physical response of the spin-Hall edge becomes also completely $\omega$ independent.
In contrast to the conventional Dicke model, a superradiant quantum phase transition occurs for arbitrarily small 
values of the coupling, $g$, while a critical coupling was required for the conventional,
homogeneous Dicke case.
Eq. \eqref{spsol1} indicates an infinite order quantum phase transition as a function of $\tilde g$ as opposed to the second order transition of the conventional Dicke model\cite{emarybrandes}.
Since $\lambda=0$ due to the lack of constraint, the photocurrent along the edge is zero.

\begin{figure}[h!]
\centering
\includegraphics[width=6.5cm]{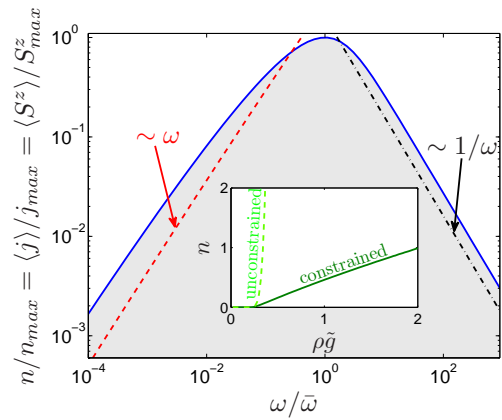}
\caption{(Color online) The weak-coupling solution of the mean field equations, Eq. \eqref{mfeqs} is plotted for the constrained case, 
the low and high frequency  asymptotes are indicates by dashed 
and dash-dotted lines, agreeing with the Floquet solution of the corresponding classical model\cite{dorafloquet}.
The unconstrained solution in Eq. \eqref{spsol1} coincides with the constrained one in the high frequency limit.
The inset shows the evolution of $n$ for $\omega\rho=1$ as a function of the Zeeman coupling
for the constrained and unconstrained cases (solid and dashed line, respectively).}
\label{nvsomega}
\end{figure}

\emph{Constrained solution.}
Now we turn to the more relevant and interesting case with a fixed total number of excitations\cite{eastham,keeling,zhangdicke} using both expression from Eq. \eqref{mfeqs}.
In the low frequency limit, the solution to Eq. \eqref{mfeqs} is obtained approximately as
\begin{gather}
n\approx \frac{(\rho g)^2}{8}\ln\left(\frac{4\sqrt 2 W}{\rho g^2}\right)=\frac{\rho^2\tilde g \omega}{8}
\ln\left(\frac{4\sqrt 2W }{\rho \tilde g\omega}\right)
\end{gather}
to logarithmic accuracy.
In the opposite, high frequency limit the solution
coincides with Eq. \eqref{spsol1}. 
Due to the conserved number of total excitations, the superradiance of the cavity mode implies a finite photocurrent, 
which is also indicated by the finite value of $\lambda$.
The relevant  energy scale of the problem separates the 
low and high frequency  limits, which is signaled by a maximum in the photon occupation number, occurring at
\begin{gather}
\bar\omega=\frac{8W}{\rho\tilde g}\exp\left(-\frac{2}{\rho\tilde g}-\frac 12\right),
\end{gather}
where the maximum of the photon number and current are
$n_{max}={\rho^2\tilde g\bar\omega}/{16}$ and $j_{max}=-{2en_{max}}/{\pi\rho}$,
as shown in Fig. \ref{nvsomega}. While the $\omega$ dependence features a maximum for fixed $\tilde g$, the
$\tilde g$ dependence of these quantities is monotonic for fixed $\omega$, shown in the inset of Fig. \ref{nvsomega}.
The photocurrent behaves  as
\begin{gather}
\langle j\rangle =\left\{\begin{array}{cc}
-\dfrac{e\omega}{2\pi}\dfrac{\rho \tilde g}{2}\ln\left(\dfrac{4\sqrt 2W}{\rho\tilde g\omega}\right) & \textmd{for } \omega\ll \bar \omega,\\
-\dfrac{8eW^2}{\pi\rho \tilde g \omega}\exp\left(-\dfrac{4}{\rho \tilde g}\right) &  \textmd{for } \omega\gg\bar\omega.
\end{array}
\right.
\end{gather}

The photocurrent arises from the constraint, which represents an effective \jav{magnetic field along the edge, inducing
 a photocurrent through the magnetoelectric effect\cite{hasankane}.}
 In the low frequency limit, the current is pseudo-quantized as it grows proportional to $\omega$ (with $\log(1/\omega)$ corrections). This is similar to the
Floquet case, where a perfectly quantized current without log-corrections follows from the adiabatic charge pumping\cite{thouless}.
The difference between the quantum and classical description arises because the condition of charge pumping that
the Hamiltonian should change slowly and periodically in time is not satisfied here by the quantized photon field and quantum fluctuations do not allow for purely
 periodic time evolution and destroy quantization.
This parallels to the breakdown of the classical equipartition theorem, which predicts vanishing energy per degree of freedom at absolute zero, 
which is violated by
quantum fluctuations, which produce the zero-point energy of a quantum harmonic oscillator.

In the high frequency limit, the current decays slowly as $1/\omega$ with increasing frequency, which again agrees with the Floquet case.
The ground state energy in this case is always above the absolute minimum energy state of the  unconstrained case, although these approach 
each other fast in the $\omega\gg\bar\omega$ limit  similarly to other quantities, because
$\lambda$ vanishes as $\sim -1/\omega$ with increasing frequency and $\lambda=0$ in the unconstrained case.
By allowing for $\mathcal Q\neq 0$ in the constrained case, the superradiant phase 
behaves similarly to $\mathcal Q=0$ and is accompanied by a finite photocurrent along the edge.
For some $\omega$ and $\tilde g$ pairs, 
the current may vanish, which indicates that the constrained solution in a given $\mathcal Q\neq 0$ sector coincides with the absolute ground state.

Due to the magnetoelectric phenomenon\cite{hasankane}, a finite magnetization develops in the $z$ direction along the 
edge as $\langle S^z\rangle=\langle j\rangle/ev$, unlike in the unconstrained case.

\begin{figure}[h!]
\centering
\includegraphics[width=3.6cm]{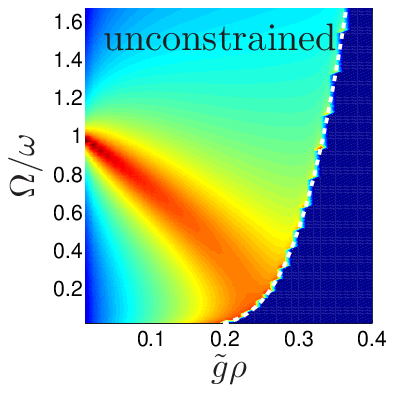}
\hspace*{2mm}
\includegraphics[width=42mm]{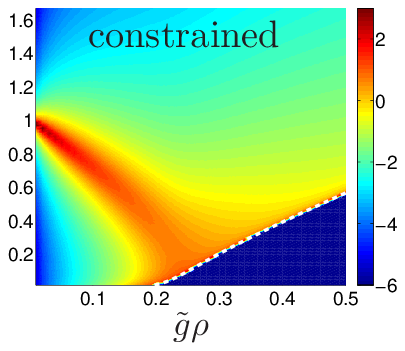}
\caption{(Color online) The contour plot of $\ln[\omega A(\Omega)]$ reveals the polariton excitation energies and their spectral weights for $W/\omega=100$, $\omega\rho=1$.
The white dashed line denotes the minimal excitation energy $2\Delta$,
above which a continuum is formed. There is always a Goldstone mode at $\Omega=0$, omitted from these plots. }
\label{photonspec}
\end{figure}

\emph{Photon spectral function.}
Having discussed the electronic properties exhaustively, now we turn to the properties  of the photon mode in the superradiant phase, which
contains the polariton spectrum and reveals the stability of the mean field solution.
Its Green's function is obtained by evaluating the Gaussian fluctuation correction of the order parameter around the mean-field value\cite{eastham}.
At $T=0$, it reveals a zero energy Goldstone mode for any finite $g$ due to the  breaking of the $U(1)$ symmetry\cite{ye}.
In the $(a,a^+)$ Nambu space, the Green's function close to zero frequency becomes singular as
\begin{gather}
\mathcal G(\Omega\approx 0)^{-1}=\left[\begin{array}{cc}
\Omega+\frac{g^2\rho}{4} & \frac{g^2\rho}{4} \\
\frac{g^2\rho}{4}        & -\Omega+\frac{g^2\rho}{4}
\end{array}\right],
\end{gather} 
and $\Omega$ is understood as $\Omega+i0^+$. The spectral function $A(\Omega)=\textmd{Im Tr}~G(\Omega)/\pi$, measurable by the absorption coefficient 
of the cavity, 
 features the Goldstone mode as
$A(\Omega\rightarrow 0)=\frac{g^2\rho}{2}\delta^\prime(\Omega)$  from the Gaussian fluctuations of the order parameter. 
The spectral weight of the Goldstone mode  vanishes at $g=0$ due to the absence of phase transition there, and the spectral function assumes its
non-interacting value as $A(\Omega\geq 0)=\delta(\Omega-\omega)$.
The structure of the spectral function for vanishing $\Omega$ agrees with that of the finite-spin Tavis Cummings model\cite{ye}.
In addition, a high energy  mode is expected as a remnant of the bare cavity mode $\omega$.
This mode gets damped as $\sim \pi g^2\rho$ for small $\Delta$ and is renormalized towards 
smaller frequencies before it hits the gap edge at $\Omega=2\Delta$ and merges with it. This optical gap is the amplitude mode (or Higgs like mode) where 
the spectral function displays a square-root singularity, as shown in Fig. \ref{photonspec}, and this sets the 
threshold energy for continuum polariton excitations for $\Omega>2\Delta$.

\emph{Discussion.}
A single photon mode in Eq. \eqref{hamilton} is realized in a quantum LC circuit\cite{todorov}  or is selected from  a ladder of cavity modes by placing a dispersive element into the cavity such as  prism or nonlinear dielectric material (with wavevector dependent refractive index).
The latter results in a non-equidistant ladder of excitations, which are eliminated by using dielectric  mirrors (with high reflectivity for the basic cavity 
mode, and high transmission for  inharmonic  frequencies). 
Tuning $\omega$ and $g$  is possible by changing
 the dielectric constant or the position of  mirrors in the cavity, thus exploring various regions of the phase diagram. 
Multimode photon fields yield  similar results as was demonstrated in Refs. \cite{wanghioe,hepp,keeling}. This is most directly seen by integrating out the photon modes,
which further enhances the effective electron-electron interaction.

\jav{The SHI is realized using either condensed matter\cite{hasankane} or cold atomic\cite{zoller,stanescu,goldman} settings.
The electromagnetic field appears not only through the Zeeman term but via a vector potential. 
Neutral cold atoms in optical traps or  chiral edge states (with spin parallel to momentum) \cite{goldman} do not couple to it, though.
Otherwise, it can be incorporated into our calculations, and its main effect is the suppression of $\omega$
in Eq. \eqref{mfeqs}, which is negligible for
$\omega\gg v ec\sqrt{\rho\tilde g}/g_{\rm eff}\mu_B$,
where $c$ is the speed of light, $g_{\rm eff}$ is the effective g-factor of the edge states and $\mu_B$ is the
Bohr magneton.}
When a weak external field is  coupled to the photon field, the phase of the order parameter can be tuned slowly and 
the quantized Chern number can be probed through a Thouless type\cite{thouless} adiabatic charge pumping.

The superradiant phase transition occurs for arbitrary small coupling between the photon field and the spin-Hall edge, and is characterized
by a Chern number $\pm 1$ in the space of the 1D Brillouin zone and the phase of the order parameter. The 
unconstrained solution, corresponding to the absolute ground state of the system, does not carry a net electric 
current. By 
keeping the total number of excitations fixed, a finite electric current carrying state develops, which behaves analogously to its Floquet counterpart.
This remotely parallels to the idea of quantum time 
crystals\cite{wilczek}, which do not occupy  the absolute ground state of a system but appear in an excited sector\cite{bruno}. The conventional Dicke model exhibits 
quantum chaos\cite{emarybrandes} away from the thermodynamic limit as a precursor to quantum phase transition at $\NN\rightarrow\infty$. It would be interesting to investigate 
whether and how quantum chaotic behaviour sets in in the present model for finite $\NN$. Extending our work to higher dimensional systems, which are suspected to become 
Floquet topological insulators such as graphene\cite{kitagawaoka} or HgTe/CdTe quantum wells\cite{lindner} or their cold atomic realizations, promises to be a fruitful 
enterprise.

\begin{acknowledgments}
We thank J. Asb\'oth, P. Domokos, F. Simon, G. Szirmai, Sz. Vajna and especially G. Zar\'and for stimulating discussions and comments, and
 acknowledge support by the Hungarian Scientific Research Fund No. K101244,  K105149, K108676,
by the ERC Grant Nr. ERC-259374-Sylo and by the Bolyai program of the 
Hungarian Academy of Sciences.
\end{acknowledgments}

\bibliographystyle{apsrev}
\bibliography{refgraph}

\end{document}